\documentclass[sigconf]{acmart}

\usepackage{subcaption}

\newcommand{\lf}{O$\leftarrow$O}
\newcommand{\rf}{O$\rightarrow$O}
\newcommand{\mf}{O$=$O}
\newcommand{\nf}{O  O}

\makeatletter
\newcommand*\bigcdot{\mathpalette\bigcdot@{.5}}
\newcommand*\bigcdot@[2]{\mathbin{\vcenter{\hbox{\scalebox{#2}{$\m@th#1\bullet$}}}}}
\makeatother

\newcommand{\mybullet}{
    \noindent
    $\rhd$
}

\copyrightyear{2021}
\acmYear{2021}
\setcopyright{iw3c2w3}
\acmConference[WWW '21]{Proceedings of the Web Conference 2021}{April 19--23, 2021}{Ljubljana, Slovenia}
\acmBooktitle{Proceedings of the Web Conference 2021 (WWW '21), April 19--23, 2021, Ljubljana, Slovenia}
\acmPrice{}
\acmDOI{10.1145/3442381.3449861}
\acmISBN{978-1-4503-8312-7/21/04}

\begin{document}

\title{The Structure of Toxic Conversations on Twitter}

\author{Martin Saveski}
\affiliation{%
  \institution{MIT}
  \country{}
}
\email{msaveski@mit.edu}

\author{Brandon Roy}
\affiliation{%
  \institution{MIT}
  \country{}
 }
\email{bcroy@media.mit.edu}

\author{Deb Roy}
\affiliation{%
  \institution{MIT}
  \country{}
}
\email{dkroy@media.mit.edu}

\begin{abstract}
Social media platforms promise to enable rich and vibrant conversations online; however, their potential is often hindered by antisocial behaviors. In this paper, we study the relationship between structure and toxicity in conversations on Twitter. We collect 1.18M conversations (58.5M tweets, 4.4M users) prompted by tweets that are posted by or mention major news outlets over one year and candidates who ran in the 2018 US midterm elections over four months.
We analyze the conversations at the individual, dyad, and group level. At the individual level, we find that toxicity is spread across many low to moderately toxic users. At the dyad level, we observe that toxic replies are more likely to come from users who do not have any social connection nor share many common friends with the poster. At the group level, we find that toxic conversations tend to have larger, wider, and deeper reply trees, but sparser follow graphs.
To test the predictive power of the conversational structure, we consider two prediction tasks. In the first prediction task, we demonstrate that the structural features can be used to predict whether the conversation will become toxic as early as the first ten replies. In the second prediction task, we show that the structural characteristics of the conversation are also predictive of whether the next reply posted by a specific user will be toxic or not. We observe that the structural and linguistic characteristics of the conversations are complementary in both prediction tasks. Our findings inform the design of healthier social media platforms and demonstrate that models based on the structural characteristics of conversations can be used to detect early signs of toxicity and potentially steer conversations in a less toxic direction.
\end{abstract}

\maketitle

\section{Introduction}
With billions of users worldwide, social media platforms have become a vital part of our lives and now constitute an important facet of our public sphere. They allow users to share their views and prompt conversations on issues that they care about. In the case of Twitter, a user can post a tweet and any other user who sees the tweet can reply, sharing their point of view, bringing the content into their network of followers, and broadening the conversation. This chain reaction of replies propagates in complex ways through the Twitter network and can lead to conversational exchanges between people who may be tightly connected, or equally to people who have never met and have little to no connection on Twitter. This potential for large-scale conversations across diverse sets of people holds promise for supporting a rich and vibrant public discourse but also permits degradation of civility between people. Antisocial behaviors online are very prevalent~\cite{vogels2021pew} and can be damaging to mental and emotional health~\cite{raskauskas2007involvement, akbulut2010cyberbullying}.

Prior work has primarily focused on characterizing and detecting various types of antisocial behaviors, including hate speech~\cite{davidson2017automated, mondal2017measurement}, harassment~\cite{founta2018large}, cyberbullying~\cite{yao2019cyberbullying}, or more generally toxic behavior~\cite{blackburn2014stfu, wulczyn2017ex}. These approaches typically analyze the content in isolation and after the fact~\cite{fortuna2018survey}. While useful for monitoring the levels of toxicity and limiting the exposure to toxic content, these methods have limited potential in forecasting toxicity and preventing toxic behaviors from occurring in the first place~\cite{jurgens2019just}.

Forecasting toxicity requires considering the social and conversational context of the discussions in which the toxic behavior occurs. Previous work on conversational analysis has examined various conversational outcomes, such as whether the conversation will grow~\cite{backstrom2013characterizing, zhang2018characterizing}, whether a participant will re-enter the conversation~\cite{backstrom2013characterizing, shugars2019keep}, whether the conversation is productive~\cite{niculae16constructive} or controversial~\cite{coletto2017automatic, hessel2019something}, and whether it will lead to disagreement~\cite{wang2014piece}. A recent line of work has focused on predicting toxicity based on the pragmatic cues~\cite{zhang2018conversations} and learned representations~\cite{chang2019trouble} of the language used in the initial exchanges of the conversation. As we show later in our prediction analyses, the linguistic characteristics of the conversations are complementary to the structural features that are the focus of this study.

Studies of the structure of conversations have mainly focused on the relations between comments, including the structural characteristics of various types of conversations~\cite{gomez2008statistical, gonzalez2010structure}, generative models of the conversation threading structure~\cite{kumar2010dynamics, gomez2011modeling, aragon2017generative}, and the effects of different conversational interface designs (e.g., hierarchical vs. linear)~\cite{aragon2017thread, budak2017threading}. However, less attention has been paid on how the structure of the conversation and the social relationships among the conversation participants affect the conversation dynamics and how they are related to toxic behaviors.

%
%
\begin{figure*}[t]
\centering
\begin{subfigure}[t]{0.2175\linewidth}
    \includegraphics[width=\linewidth]{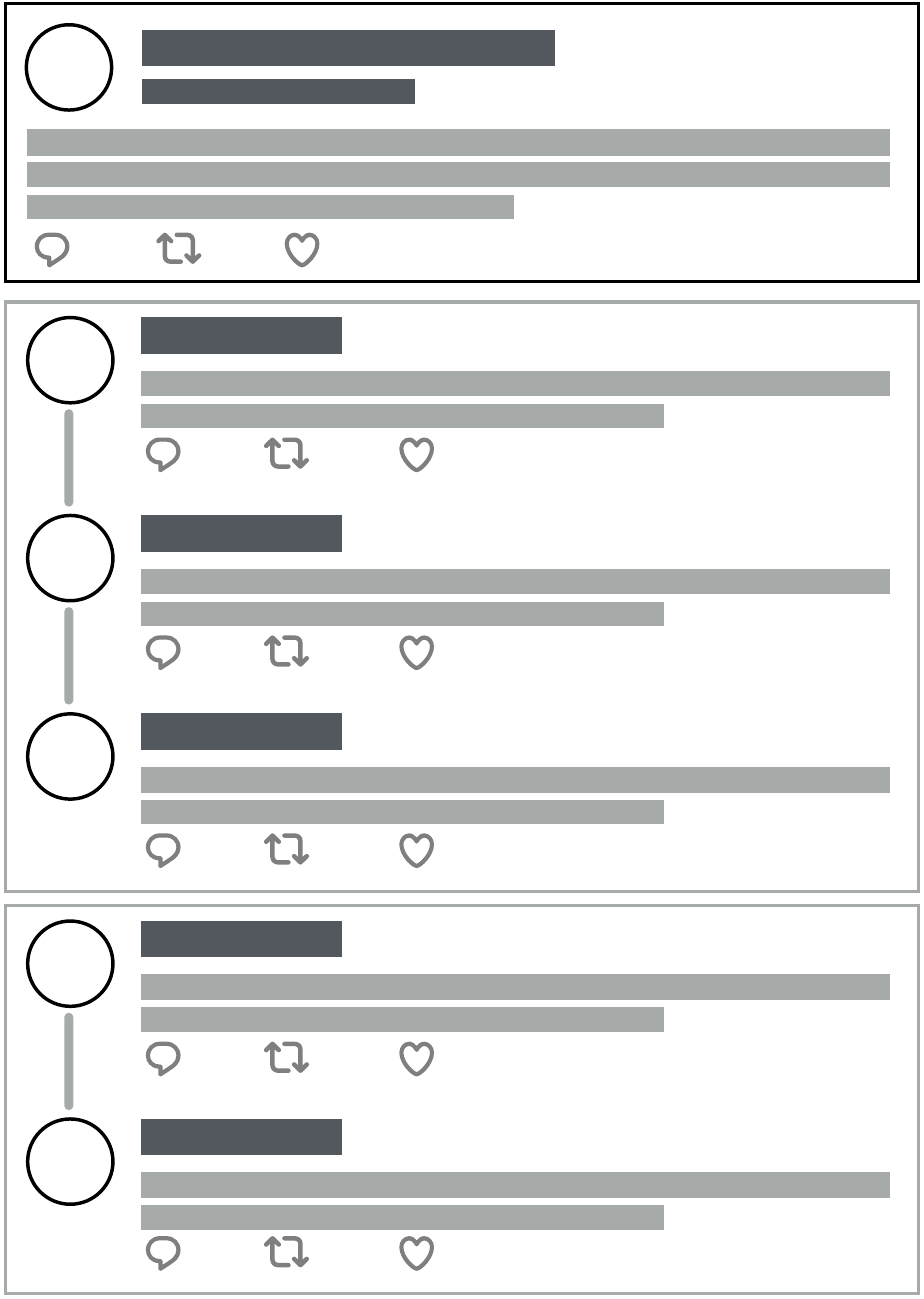}
    \caption{Twitter User Interface}
\end{subfigure}
\hfill
\begin{subfigure}[t]{0.245\linewidth}
    \includegraphics[width=\linewidth]{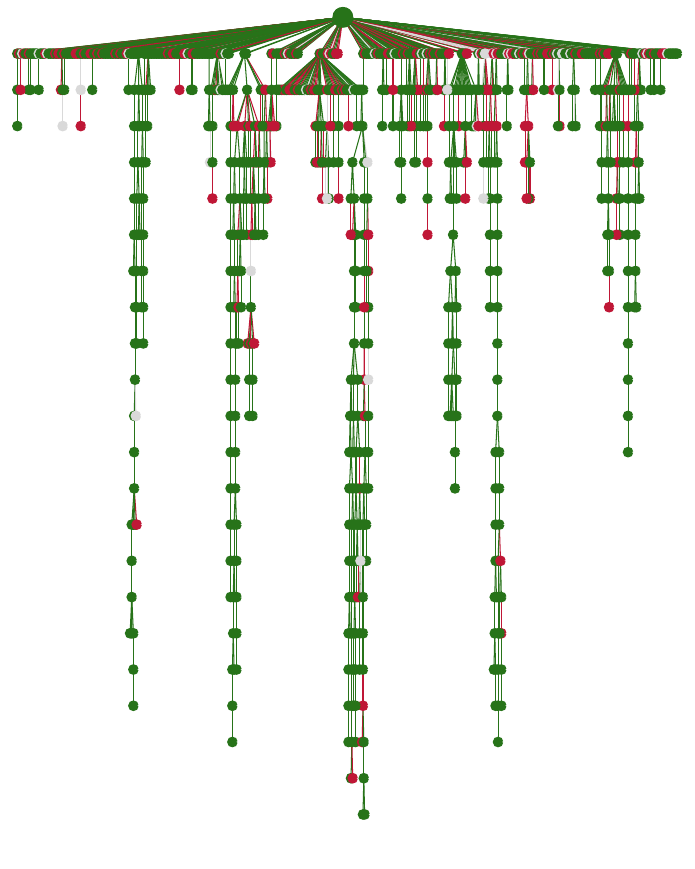}
    \caption{Reply Tree}
\end{subfigure}
\hfill
\begin{subfigure}[t]{0.226\linewidth}
    \includegraphics[width=\linewidth]{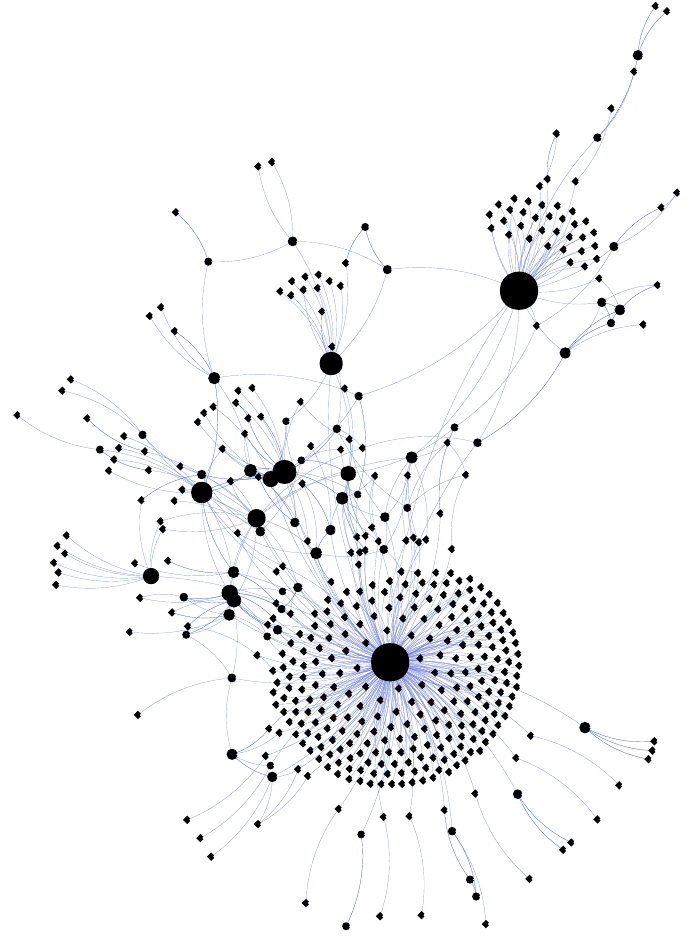}
    \caption{Reply Graph}
\end{subfigure}
\hfill
\begin{subfigure}[t]{0.226\linewidth}
    \includegraphics[width=\linewidth]{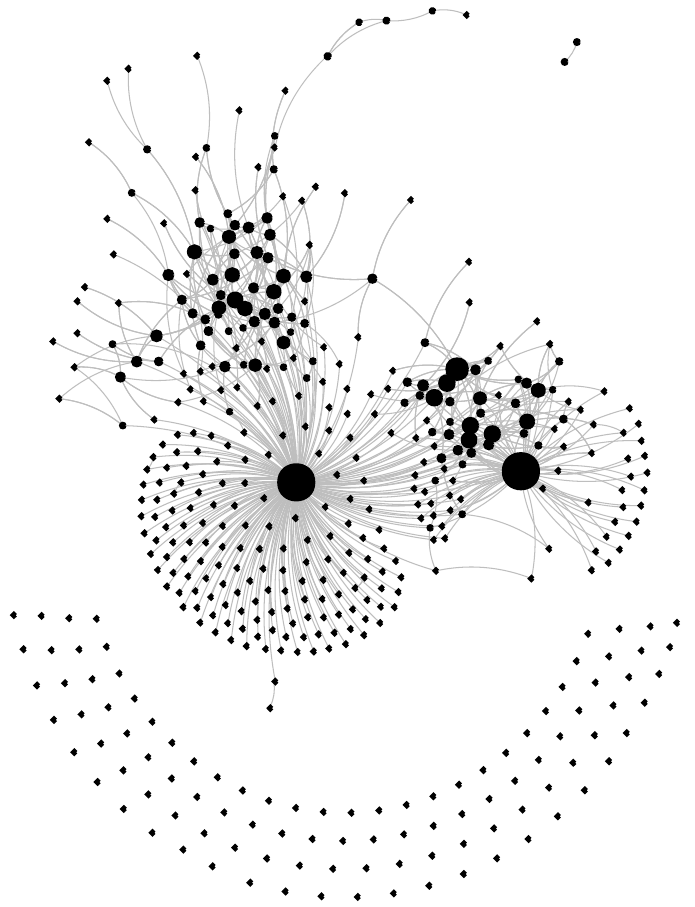}
    \caption{Follow Graph}
\end{subfigure}
\caption{Four views of a Twitter conversation started by a @foxnews tweet. (A) A sketch of the conversation as experienced by the conversation participants through the Twitter UI. (B) Reply tree, the root node is the tweet that prompted the conversation and the remaining nodes are replies. The red nodes represent tweets classified as toxic. (C) Reply graph, a user-centric view of the reply tree in which two users are connected if one replied to the other, and (D) the graph of the follow relationships between the conversation participants. The size of the nodes in C and D is proportional to their PageRank.}
\label{fig:dtox-hero}
\end{figure*}

\textbf{The present work. }
The starting point for the present work is the observation that communication is a social act and that the relationships between the conversation participants may influence their behaviors. To systematically investigate the relationship between structure and toxicity in conversations on Twitter, we collected a sample of 1.18M conversations (58.5M tweets, 4.4M users) prompted by tweets that are posted by or mention five major news outlets (CNN, New York Times, Wall Street Journal, Fox News, and Breitbart) and 1,430 candidates who ran for office in the 2018 US midterm elections\footnote{The code and (anonymized) data needed to replicate our analyses are available at: \url{https://github.com/msaveski/toxic_conversation_structure/}}.

We represent the structure of a conversation in three ways: using a \textit{reply tree} (Figure~\ref{fig:dtox-hero}B) which encodes the relationships between posts, where two posts are connected if one is a reply to the other; a \textit{reply graph} (Figure~\ref{fig:dtox-hero}C), a directed graph which captures the conversational interactions between users, where two users are connected if one replied to the other; and a \textit{follow graph} (Figure~\ref{fig:dtox-hero}D), which captures the social connections among the conversation participants, where one user is connected to another if they follow~them.

The goal of this work is twofold: ($i$) to study the relationship between the conversational structure and toxicity, and ($ii$) to test the value of the structural view of the conversations in forecasting future toxicity as the conversation unfolds. To study the link between structure and toxicity, we analyze the conversations at three levels: individual, dyad, and group level. To measure the predictive power of the structural characteristics of the conversations, we consider two prediction tasks. In the first task, we predict whether the conversation will become more or less toxic, based on the initial stages of the conversation. In the second task, we predict whether the next reply posted by a specific user will be toxic, given the conversation so far and the user’s relationship with the conversation~participants.

In our analysis of the conversations, we find that, at the individual level, toxicity is not concentrated among a small number of highly toxic users but dispersed among many low to moderately toxic users (\S \ref{subsec:individual-level}). We also observe a moderate level of homophily among users who posted at least one (or at least four) toxic tweet(s). At the dyad level, we find that toxic posts are more likely to elicit toxic replies than nontoxic posts (\S \ref{sec:dyads}). Toxic replies are more likely to come from other users who do not have any social relationship with the poster, do not have many common friends, and have fewer followers. At the group level, we find that toxic conversations tend to have larger, wider, and deeper reply trees (\S \ref{sec:analyses-reply-tree}). However, conversations in which the follow graph among the conversation participants is denser, has fewer connected components, and lower modularity tend to be less toxic (\S \ref{sec:analyses-follow-graph}).

In the first prediction task, we find that we can predict whether a conversation will become more or less toxic with an accuracy of 61.6\% (AUC:~66.2\%) in the news and 59.9\% (AUC:~64\%) in the midterms dataset given the first ten replies, using only the structural characteristics of the conversation and after controlling for the toxicity in the initial ten replies (\S \ref{sec:prefix-predictions}). In the second prediction task, we find that we can predict whether the next reply posted by a specific user will be toxic with an accuracy of 68\% (AUC:~75.3\%) in the news and 70.5\% (AUC:~79.9\%) in the midterms dataset, even after controlling for the content of the tweet that prompted the conversation (\S \ref{sec:next-reply-predictions}). In both prediction tasks, we observe that combining the structural features with features that encode the linguistic characteristics of the conversation further improves the classification performance, suggesting that the two types of features capture different and complementary aspects of the conversations.

These results suggest changes in the design of social media platforms that could reduce toxicity at scale.

%
%
\section{Data} \label{sec:conv-data}

\textbf{Account selection.}
To capture a wide variety of conversations, we collected conversations prompted by major news outlets and candidates who ran for office during the 2018 midterm elections in the US. We selected five news outlets that span the political spectrum---New York Times and CNN on the left, Wall Street Journal in the middle, and Fox and Breitbart on the right~\cite{bakshy2015exposure, budak2016fair}---and have Twitter accounts with a large number of followers. We collected both the conversations started by tweets posted by these accounts and by tweets posted by others that @mention these accounts.

We tracked the news accounts for one year, from May 2018 to May 2019, capturing 510k conversations (32.6M tweets, 2.4M users), and the accounts of the midterm candidates for five months, one month leading up to the election and four months after, capturing 676.8k conversations (25.8M tweets, 2M users). We followed both the personal accounts that the candidates used during their campaigns and their official accounts created after their inauguration. We obtained the personal Twitter accounts of the candidates from \textit{Ballotpedia}, and the official accounts from the \textit{congress-legislators} Github repository. 1,430 of 3,339 candidates had a Twitter account.

Taken together, the two datasets include a large number of conversations over a long period of time. Moreover, the collected conversations vary in several important ways. They capture discussions prompted by a politically diverse set of accounts, including both left- and right-leaning news outlets and midterm candidates. Some conversations are started by highly influential accounts such as the news outlets and the candidates with a large number of followers, others by ordinary users who @mentioned the news outlets or the candidates in their tweets.

\textbf{Data collection pipeline.}
The key technical challenge in collecting tweets related to the same conversation is that the Twitter API only provides a link from the reply to the original tweet, but not vice versa\footnote{We note that after we collected the data, Twitter introduced a new API endpoint that allows conversation threads to be easily retrieved.}.
Thus, given a root tweet, one cannot simply query for all subsequent replies. To overcome this issue, we rely on the fact that every time a user replies to a tweet, they implicitly @mention all users that posted or were mentioned earlier in the reply chain. We use the Twitter PowerTrack API to collect all posts and mentions of the selected accounts over the study period.
To string together the replies and build the complete reply trees (Figure~\ref{fig:dtox-hero}B), we scan the full dataset and use the \textit{reply-to} field to recursively link posts to replies. We retain only reply trees rooted in tweets that are either posted by or @mention the selected accounts and exclude tweets with no replies or strings of replies by only one user.

To collect the social graphs of the users who participated in these conversations, we set up a daily job that scans all tweets collected in the last 24 hours, compiles a list of all users that posted at least one tweet, and using the Twitter REST API downloads each user's list of friends and followers. We can thus use the user's follow graph snapshot corresponding to the time of their tweet. We do not collect data on accounts that are protected.

\section{Toxicity Annotations} \label{sec:toxicity-annotations}
To label tweets for toxicity, we used Google's Perspective API~\cite{wulczyn2017ex}. We chose this API as its models are trained on Wikipedia comments, which like tweets, are short and informal. The initial Perspective API model was trained on 100k comments each annotated 10 times and was reported to be as accurate as the aggregate performance of three annotators. Since then, the model has been retrained on a larger dataset and modified to address some of its weaknesses reported by other researchers (e.g.,~\cite{sap2019risk}). Several other studies have used the Perspective API and have demonstrated that its predictions are accurate~\cite{rajadesingan2020quick, hua2020characterizing}.

Since the rest of our analysis relies on the Perspective API’s toxicity annotations, we thoroughly assess their quality. To do so, we deployed an Amazon Mechanical Turk annotation task to obtain human toxicity labels on randomly selected tweets. Beyond assessing the quality of the annotations, we also relied on the human annotations to tune the Perspective API score threshold that we used for classifying a tweet as toxic vs. nontoxic. (The API returns an estimate of the probability that a reader would perceive the comment as toxic, rather than a binary toxicity label.)

The Mechanical Turk annotation task consisted of five randomly selected tweets. We showed an input label next to each tweet for the annotators to select between ``toxic'' and ``nontoxic.'' To avoid any annotation bias due to ordering effects, we randomized the order of the labels between tasks (i.e., batches of five tweets), but kept the order consistent within a task. To help clarify what constituted a toxic tweet, we provided the annotators with simple instructions. We used the same definition of toxicity as the Perspective API: ``a rude, disrespectful, or unreasonable comment that may make you leave a discussion''~\cite{wulczyn2017ex}. To ensure the quality of the labels, we recruited only annotators from the US with high performance on previous Mechanical Turk tasks. We compensated them 20 cents per task (i.e., labeling five tweets). Before the annotators started the task, we warned them that they might see offensive content. The protocol was approved by the MIT institutional review board.

We randomly sampled 3,000 tweets for annotation from the first five months of the news dataset. We ensured that the sample is representative of the overall distribution of toxicity scores, as predicted by the Perspective API (K-S test, $D$ = 0.01, $p$ = 0.89). Each tweet was independently labeled by three different workers so that we can measure the inter-annotator agreement and use a voting scheme to obtain a single ``ground-truth'' label. To assess the inter-annotator agreement, we used Krippendorff's $\alpha$~\cite{krippendorff2011computing} and found a fair agreement between the annotations, $\alpha$ = 0.32. To obtain a single label for each tweet, we used a majority vote.

We tuned the Perspective API toxicity score threshold above which we consider a tweet to be toxic, and measured the quality of the predictions. We used a random sample of 600 annotated tweets (20\%) as a development set on which we chose the threshold, and the remaining tweets as a test set. We picked a threshold (T = 0.531) that strikes a balance between precision and recall on the development set. On the test set, this threshold yields a classification accuracy of 0.82, AUC of 0.86, and an F1 score of 0.63. When we consider only the subset of the test set in which annotators reached a consensus, all measures of the prediction performance increase significantly, accuracy: 0.91, AUC: 0.95 , F1:~0.73.
We note that more conservative toxicity thresholds (T~=~0.6 or T~=~0.7) lead to the same patterns in all subsequent analyses.

%
%
\section{Analyses} \label{sec:analyses}
In this section, we investigate the relationship between the conversation structure and toxic behavior at multiple scales. First, we study individual users’ propensity for toxic behavior. Second, we investigate the dyadic relationship between users by considering pairs of tweets and replies. Finally, we look at the overall conversation structure, including the reply tree and follow graph structure. To improve the readability of the text, we communicate the uncertainty of our point estimates graphically and show 95\% confidence intervals in the subsequent figures.

\subsection{Individual Level} \label{subsec:individual-level}
\begin{figure}[t]
\centering
\includegraphics[width=\linewidth]{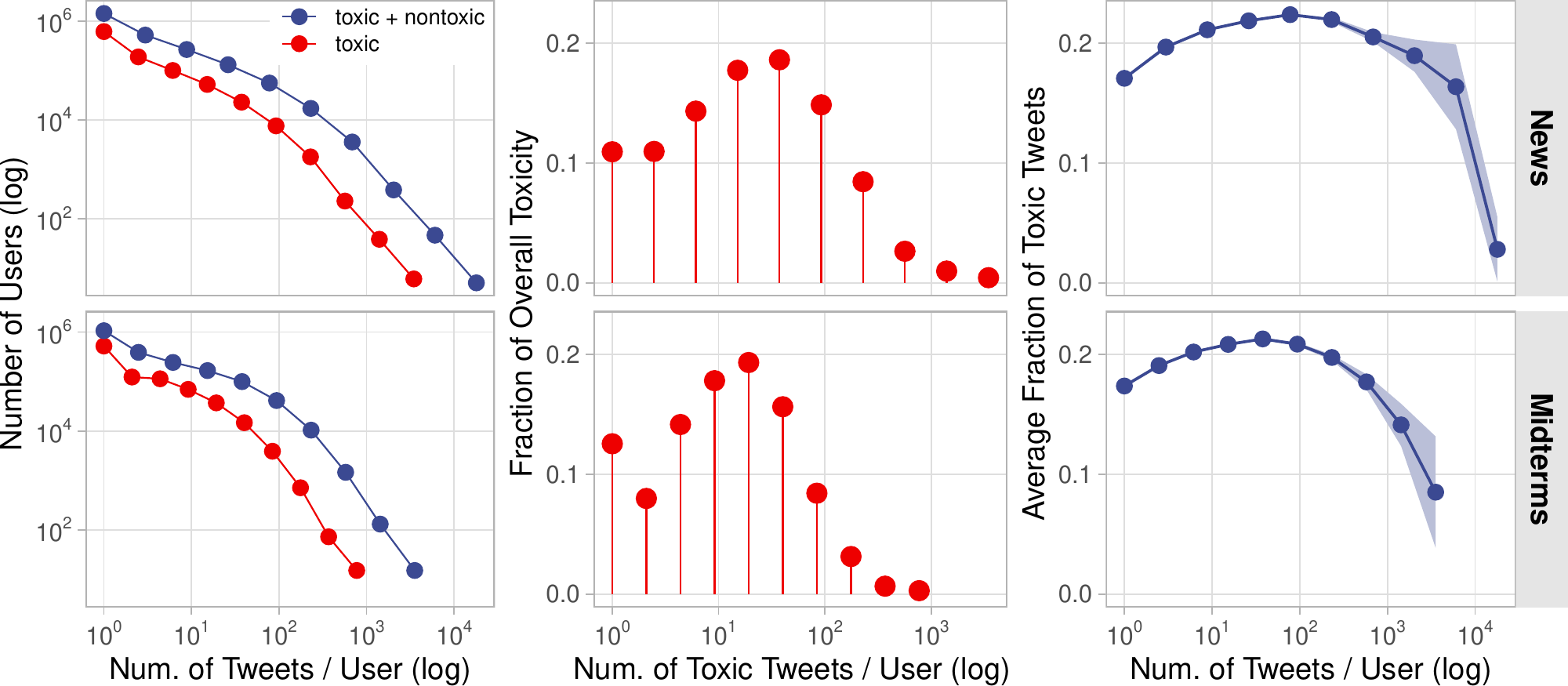}
\caption{Distribution of the number of tweets and toxic tweets per user (left). Fraction of overall toxicity contributed by users with different levels of toxicity (middle). Average fraction of toxic tweets by users with different activity levels (right). The error bands represent 95\% CIs.}
\label{fig:analysis-ego}
\end{figure}

We start by analyzing the distribution of tweets and toxic tweets per user in the two datasets. In Figure~\ref{fig:analysis-ego} (left), we bucket users in logarithmically-sized buckets by the number of tweets and toxic tweets they posted ($x$ axis) and show the number of users that fall into each bucket ($y$ axis). As one may expect, we find that both distributions are long-tailed, i.e., there are many users who posted a few tweets and a few users who posted many tweets. Out of all users, 44.71\% in the news and 38.85\% in the midterms dataset posted only one tweet. Most users---59.26\% in the news and 56.15\% in the midterms dataset---did not post any toxic tweets.

\textbf{Distribution of toxicity.}
Next, we look at how the overall toxicity is spread across the users that posted at least one toxic tweet. In particular, we are interested in whether the toxicity is concentrated among a small number of users or dispersed across the population. This has important implications on how the platform might approach reducing toxic behavior. For instance, if only a small fraction of users are toxic, one may hope that changing their behavior or altogether removing them from the platform may disproportionately reduce the overall toxicity and significantly improve the experience for the rest of the users on the platform.

In Figure~\ref{fig:analysis-ego} (middle), we bucket users in logarithmically-sized buckets by the number of toxic tweets they posted and compute what fraction of toxic tweets (out of all toxic tweets in the dataset) was posted by users in each bucket. We find a very similar pattern in the two datasets: buckets containing moderately toxic users account for the largest fraction of toxic tweets, ranging from 15\% to 18\% per bucket. While there are more users in the lower toxicity buckets, the higher number of toxic tweets per user in the medium toxicity buckets leads to a larger number of toxic tweets. This suggests that the toxicity is not concentrated among a few highly-toxic users, but it is rather dispersed across many low to moderately toxic users.

\textbf{Rate of toxicity.}
Next, we investigate how the fraction of toxic tweets per user varies for users with different activity levels. In Figure~\ref{fig:analysis-ego} (right), we bucket users in logarithmically-sized buckets by the number of tweets they posted and measure how often their tweets were toxic. We find a similar pattern in both datasets: moderately active users have a higher fraction of toxic tweets than both low- and high-activity users. We also find that highly active users have, on average, a smaller fraction of toxic tweets than the low activity users. We note that the estimates for the buckets of highly active users have wider confidence intervals as fewer users belong to these buckets.

%
%

\textbf{Homophily.}
We test whether there is \textit{homophily}~\cite{mcpherson2001birds} among users within the Twitter follow graph, i.e., whether toxic users are more likely to follow other toxic users and whether nontoxic users are more likely to follow other nontoxic users. We note that we are only interested in measuring homophily and do not intend to discern between homophily and social influence.

To construct the complete follow graph among the users, we use the earliest snapshot of each user's friends. To measure the levels of homophily, we use the \textit{assortativity coefficient} defined in \cite{newman2003mixing} which quantifies whether nodes with the same attributes connect more or less often than we would expect by chance, i.e., in a random network. The assortativity coefficient can take values between -1: perfect disassortativity (i.e., users connect only with others different from them) and 1: perfect assortativity (i.e., users connect only with others like them).

We start by assigning users to two categories: ($i$) users that did not post any toxic tweets and ($ii$) users that posted at least one toxic tweet, and compute the corresponding assortativity coefficient. We find that there is a moderate level of homophily among the users in the two datasets, 0.15 in the news and 0.125 in the midterms dataset. If we consider only users that did not post any toxic tweets and users that posted at least four toxic tweets, such that we exclude cases where users may be in the toxic category because a few of their tweets were misclassified, the assortativity coefficient increases to 0.228 in the news and 0.2 in the midterms dataset.

We also compute the assortativity coefficient among the users using the number of toxic tweets as an attribute. This allows us to test whether users with many toxic tweets tend to follow other users with many toxic tweets. The resulting assortativity coefficients are very close to zero, 0.006 in the news and 0.034 in the midterms dataset. If we restrict the analysis only to users with at least one toxic tweet, the assortativity coefficients are even closer to zero. These results suggest that there is neither a positive nor negative affinity for highly toxic users to connect to other highly toxic users.

In summary, we find a moderate level of homophily among users with no toxic tweets and users with at least one or at least four toxic tweets. However, we find no evidence that highly toxic users are more likely to connect to other highly toxic users.

\subsection{Dyads}\label{sec:dyads}
Next, we focus on the relationship between toxicity and the characteristics of the reply dyads. A reply dyad $(i, j)$ consists of two conversation participants, user $i$ and user $j$, where user $j$ replied to user $i$'s tweet. We call user $i$ a parent (or a poster) and user $j$ a child (or replier), since $i$'s tweet is a parent of $j$'s tweet in the reply tree. Note that user $i$ might be a child in another dyad, e.g., $(x,i)$, or user $j$ might be a parent in a dyad $(j,y)$ (e.g., if a reply tree has a branch $(x, i, j, y)$). We exclude reply dyads that are self-replies or direct replies to the root tweet as we are interested in understanding the relationship between the conversation participants. After filtering, we end up with a total of 9.2M dyads in the news and 8M dyads in the midterms dataset.

\textbf{Dyad characteristics.}
We define four dyad characteristics: ($i$) toxicity type, ($ii$) edge type, ($iii$) influence gap, and ($iv$) embeddedness.
Each dyad can be characterized by whether the parent's post is toxic and whether the child's reply is toxic, leading to four possible \textit{toxicity types}. The dyad can also have one of four \textit{edge types} depending on the relationship between the dyad users in the follow graph: ($i$) they may mutually follow each other (\mf), ($ii$) the child (replier) may follow the parent but not vice versa (\lf), ($ii$) the parent may follow the child (\rf), and ($iv$) they may not be connected at all (\nf). (Note that, in our notation, the parent user is always on the left.) The dyad's \textit{influence gap} is the ratio between the parent's and the child's number of followers. Finally, the \textit{dyad embeddedness} measures the extent to which the social contexts of the dyad users overlap. We define it as the number of common friends between the dyad users, i.e., the number of users that both follow.

%
%
\begin{figure}[t]
\centering
\includegraphics[width=\linewidth]{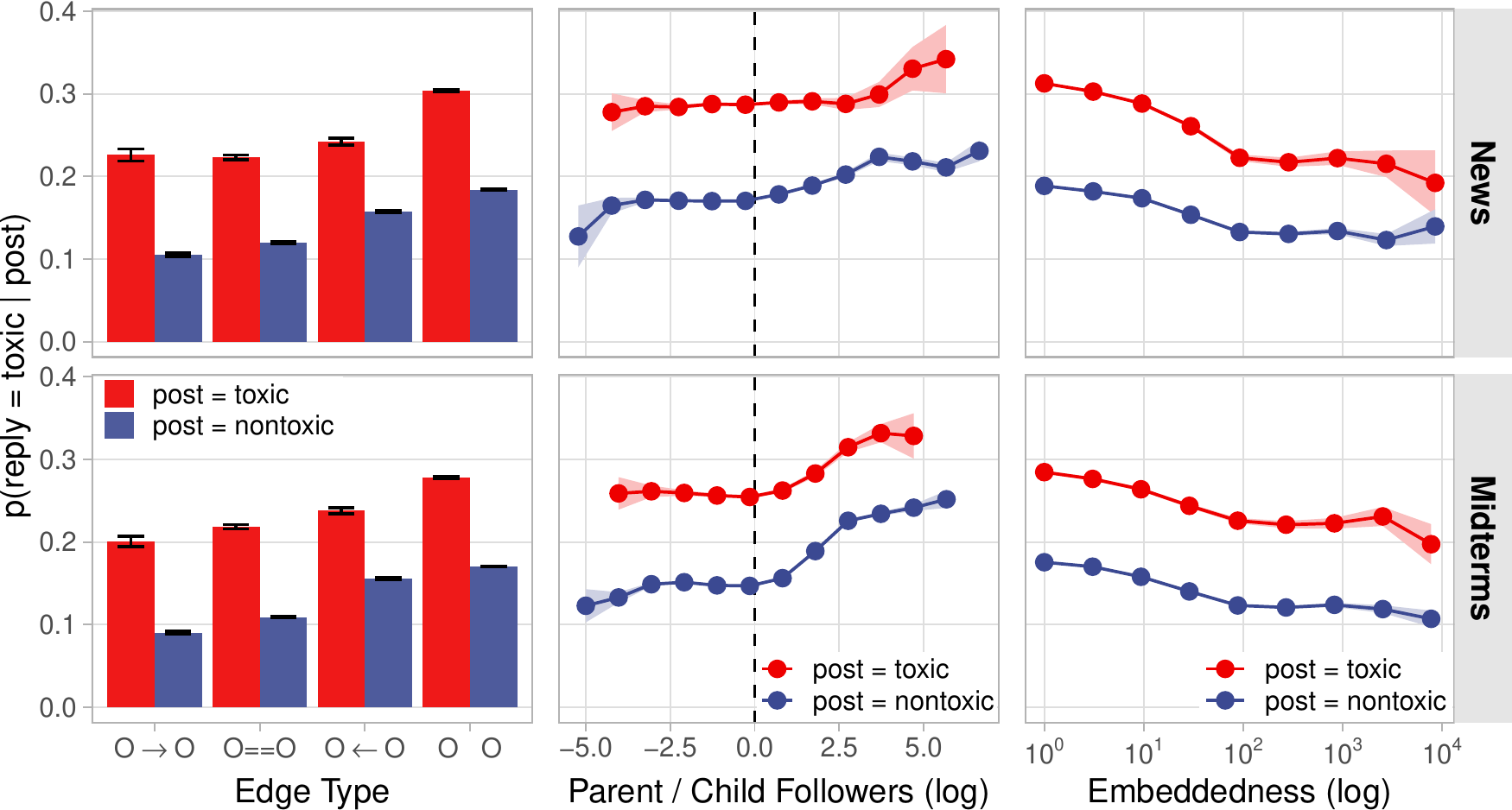}
\caption{Probability of a toxic reply given a toxic or nontoxic post depending on the edge type, the influence gap ($\log_{10}$(|parent's followers|) - $\log_{10}$(|child's followers|)), and the embeddedness (number of common friends).}
\label{fig:analysis-dyads}
\end{figure}

\textbf{Toxicity type.}
We start by analyzing how the probability of a toxic reply varies depending on whether the parent post is toxic or not. We find that toxic tweets are 65\% and 64\% more likely than nontoxic tweets to elicit toxic replies in the news and midterms datasets, respectively. The probability of a toxic reply given a toxic post is 0.3 in the news and 0.27 in the midterms dataset, while the probability of a toxic reply given a nontoxic post is 0.18 in the news and 0.16 in the midterms dataset. The toxicity type is the most defining characteristic of the dyad. We find that the patterns in other dyad characteristics differ significantly depending on whether the parent post is toxic or not. Therefore, in all subsequent analyses, we report how our findings differ in these two cases.

\textbf{Edge type.}
Next, we look at how toxicity varies across different edge types. We find that the probability of a toxic reply varies significantly depending on the edge type (Figure~\ref{fig:analysis-dyads}). Given a toxic post, a toxic reply is more likely to come from another user who neither follows nor is followed by the parent user (news: 0.30, midterms: 0.28). The probability of a toxic reply among the other edge types (\mf, \lf, or \rf) is similar, ranging between 0.22 and 0.24 in the news and between 0.2 and 0.24 in the midterms~dataset.

Given a nontoxic post, it is more likely that a toxic reply will be posted by another user who does not have any follow relationship with the poster (news: 0.18, midterms: 0.17). However, in this case, the probability that a toxic reply comes from a user who follows the poster, but not vice versa (\lf), is higher (news: 0.158, midterms: 0.156) compared to the other two edge types (\mf, news: 0.12, midterms: 0.11; or \rf, news: 0.10, midterms: 0.09). This suggests that more influential users are more likely to be a target of toxic replies. We investigate this hypothesis next.

\textbf{Influence gap.}
We define the influence gap as the ratio between the parent's and the child's number of followers. Since the distribution of the number of followers is long-tailed, we compute the log of the ratio: $\log_{10}$(|parent's followers|) - $\log_{10}$(|child's followers|). Although most dyadic interactions occur among users with a similar number of followers, users are more likely to reply to tweets posted by others who have more followers than they do. In the news dataset, when the parent's post is toxic, the probability of a toxic reply is roughly the same, regardless of the influence gap (Figure~\ref{fig:analysis-dyads}, middle). In contrast, in the midterms dataset, the probability of a toxic reply increases when the parent has more followers than the child. When the parent's post is nontoxic, then the influence gap matters even more. In both datasets, the probability that a reply will be toxic is higher when the parent has significantly more followers than the child. Interestingly, this relationship is asymmetric, i.e., the probability of a toxic reply does not decrease when the child has more followers than the parent. We find that the effect of the influence gap is most pronounced among dyads where the two users do not have any follow relationship (\nf) and when the replier follows the poster but not vice versa (\lf).

\textbf{Embeddedness.}
We define the embeddedness of a dyad as the number of common friends between the poster and the replier. Higher embeddedness suggests that the two users have similar interests and overlapping social contexts. This may influence the behavior of the replier: their potentially toxic behavior is more likely to be observed by others that both the poster and the replier are aware of and may increase the social cost of toxic behavior~\cite{coleman1988social}.

We find that the probability of a toxic reply significantly decreases as embeddedness increases (Figure~\ref{fig:analysis-dyads}, right), regardless of whether the parent post is toxic or not. For a toxic post, the probability of a toxic reply is 11\% lower in the news (dropping from 0.315 to 0.206) and 9\% lower in the midterms dataset (from 0.29 to 0.2) if the poster and the replier have 100 vs. 1 common friend. Similarly, for a nontoxic post the probability of a toxic reply decreases from 0.191 to 0.134 in the news and from 0.178 to 0.123 in the midterms dataset when the dyad users have 100 vs. 1 common friend. Like the influence gap, the effect of embeddedness is most pronounced among dyads where the two users do not have any follow relationship (\nf) and, given a nontoxic post, also among dyads in which only the replier follows the poster (\lf).

\subsection{Reply Tree Structure}\label{sec:analyses-reply-tree}
When a user posts a tweet, other users may post a reply tweet, which in turn can lead to subsequent replies. The result is a reply tree rooted in the original tweet (Figure~\ref{fig:dtox-hero}B). Here, we investigate the relationship between the structural characteristics of the reply tree and the overall toxicity of the conversation. We define the toxicity of a reply tree as the fraction of toxic tweets. The results presented are also consistent with a slightly different definition that uses the mean or the median of the toxicity scores.

\begin{figure}
\centering
\includegraphics[width=\linewidth]{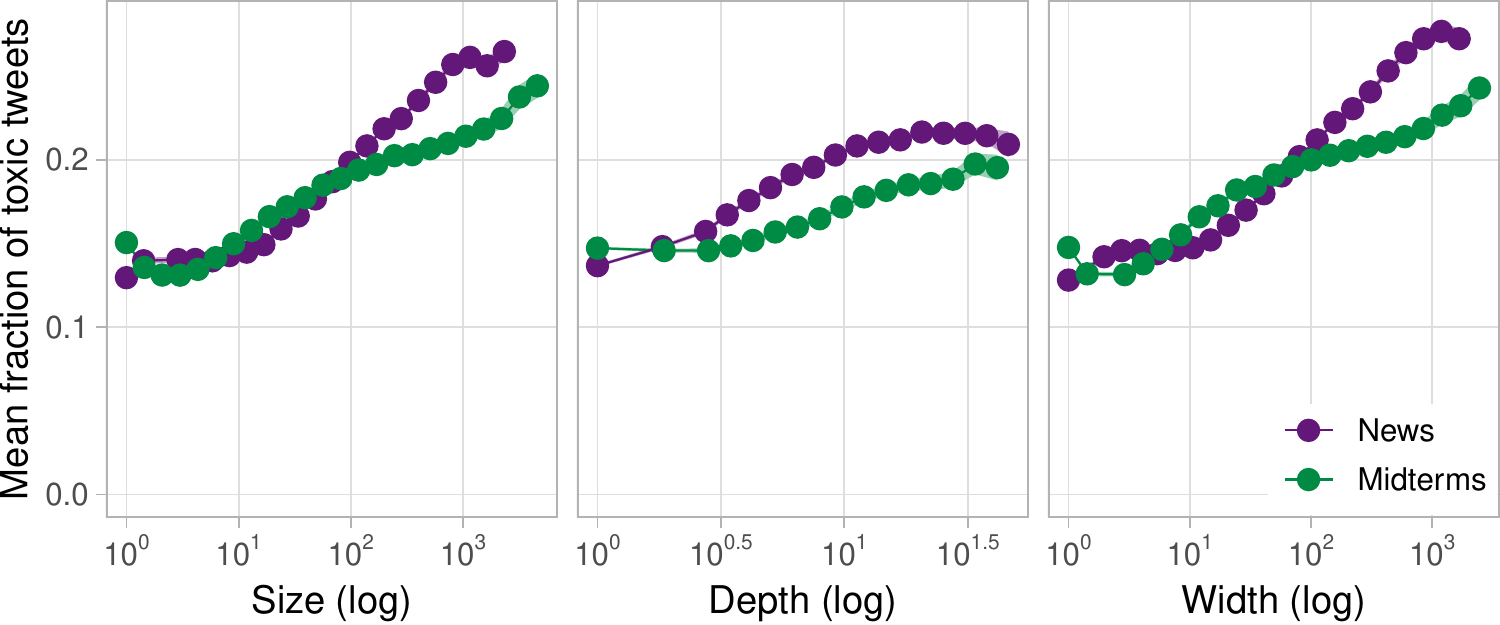}
\caption{Relationship between the size (number of tweets), depth, and width of the reply tree and the level of toxicity in the conversation. The error bands are 95\% CIs.}
\label{fig:analysis-reply-tree-1}
\end{figure}

\textbf{Size.}
First, we consider the size of the reply tree, i.e., the number of tweets in the conversation. We find a clear positive relationship between size and toxicity. Larger trees tend to be more toxic both in the news and the midterms dataset (Figure~\ref{fig:analysis-reply-tree-1}, left). We find similar results if we define size as the number of users in the conversation.

\textbf{Depth and width.}
Next, we consider the depth and the width of the reply trees. We define the tree depth as the depth of its deepest node, and tree width as the maximum number of nodes at any depth in the tree. We find that conversations with wider and deeper reply trees tend to be more toxic in both datasets (Figure~\ref{fig:analysis-reply-tree-1}, middle and right). We note that both metrics are positively correlated with tree size (news: $r_{depth}$ = 0.53,  $r_{width}$ = 0.97; midterms: $r_{depth}$ = 0.48, $r_{width}$ = 0.97) and may be proxies for size. In \S \ref{sec:prefix-predictions}, we will evaluate their usefulness as features in a predictive task.

\textbf{Wiener index.}
We investigate the Wiener index, a metric that helps us characterize the internal structure and complexity of the reply trees. The Wiener index $w(T)$ of a reply tree $T$ is defined as the average distance between all pairs of nodes:
\[ w(T) = \frac{1}{n(n-1)} \sum_{i=1}^n \sum_{j=1}^n d_{ij}, \]
where $d_{ij}$ denotes the length of the shortest path between nodes $i$ and $j$. The Wiener index was initially proposed in mathematical chemistry to characterize the structure of a molecule~\cite{wiener1947structural}. More recently, it has been used to characterize the structure of information diffusion cascades, and in particular to quantify whether information spreads in broadcast or viral fashion~\cite{goel2015structural}.
The Wiener index interpolates between two extremes: reply trees in which participants only respond to the original tweet and do not engage with each other (low $w(T)$), and reply trees with a single branch in which participants have many back-and-forth exchanges (high~$w(T)$).

\begin{figure}
\centering
\includegraphics[width=\linewidth]{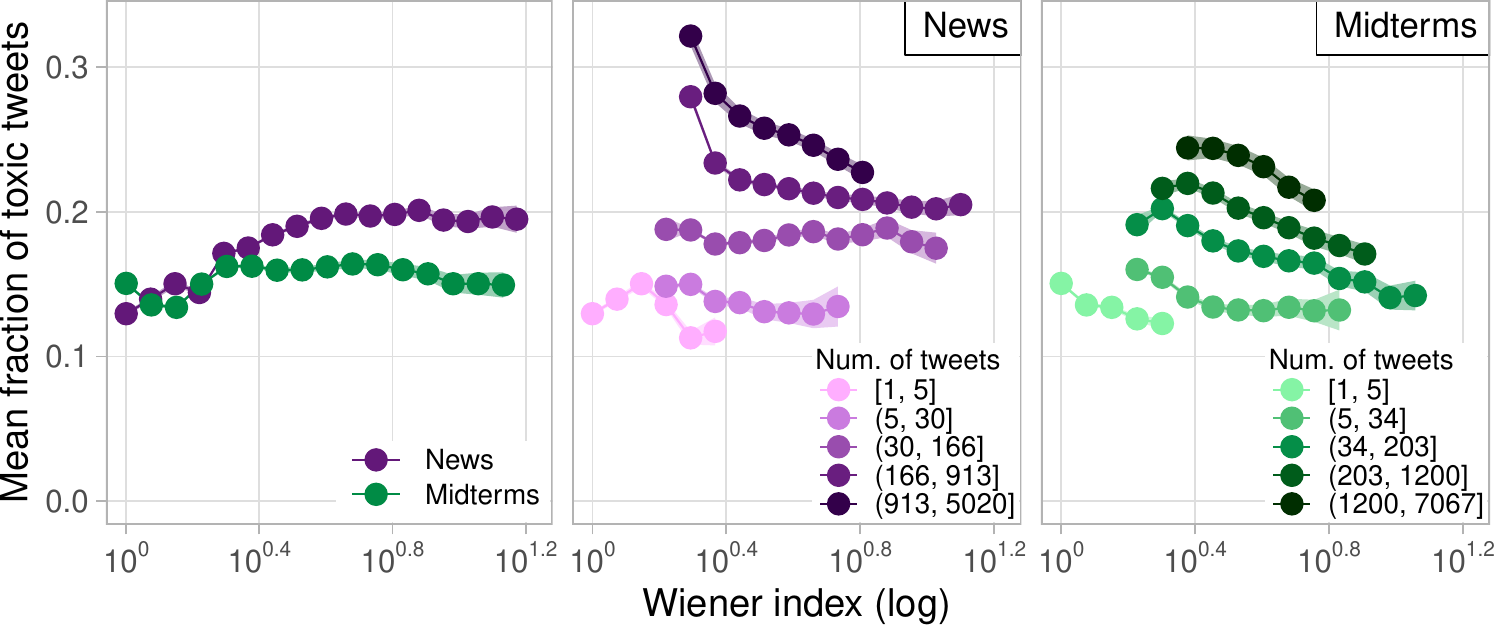}
\caption{Relationship between the reply tree Wiener index and the mean fraction of toxic tweets, in overall (left) and by tree size in the news (middle) and midterms (right) datasets.}

\label{fig:analysis-reply-tree-2}
\end{figure}

In the news dataset, we find that reply trees with a larger Wiener index tend to be more toxic; while in the midterms dataset, the mean toxicity of reply trees with varying Wiener index is mostly the same, except for a small fluctuation for trees with a low Wiener index (Figure~\ref{fig:analysis-reply-tree-2}, left).

A more complicated picture emerges when we plot the relationship between the Wiener index and toxicity for reply trees of different sizes. In Figure~\ref{fig:analysis-reply-tree-2} (middle and right), we divide all reply trees into five logarithmically-sized groups according to their size. We chose the largest number of groups that will leave us with enough data points to compare the relationship between the Wiener index and toxicity. In the news dataset, we find that for smaller reply trees, the toxicity of the conversations does not vary as a function of the Wiener index; however, for larger reply trees, we find that the toxicity decreases as the Wiener index increases (Figure~\ref{fig:analysis-reply-tree-2}, middle). In the midterms dataset, the fraction of toxic tweets in the conversation decreases as the Wiener index increases for all tree sizes, although the negative correlation is stronger for larger trees (Figure~\ref{fig:analysis-reply-tree-2}, right). Regression analysis confirms this negative relationship between toxicity and Wiener index when controlling for the number of tweets.

%
%
\subsection{Follow Graph Structure} \label{sec:analyses-follow-graph}
Next, we investigate the relationship between the structure of the follow graph (Figure~\ref{fig:dtox-hero}D) among the conversation participants and the overall toxicity of the conversations. As before, we define the overall toxicity of the conversation as the fraction of toxic tweets. We note that the conversation participants have only a local view of the follow graph; they may recognize their friends or followers, but are unlikely to know how other participants are connected.

\begin{figure}
\centering
\includegraphics[width=\linewidth]{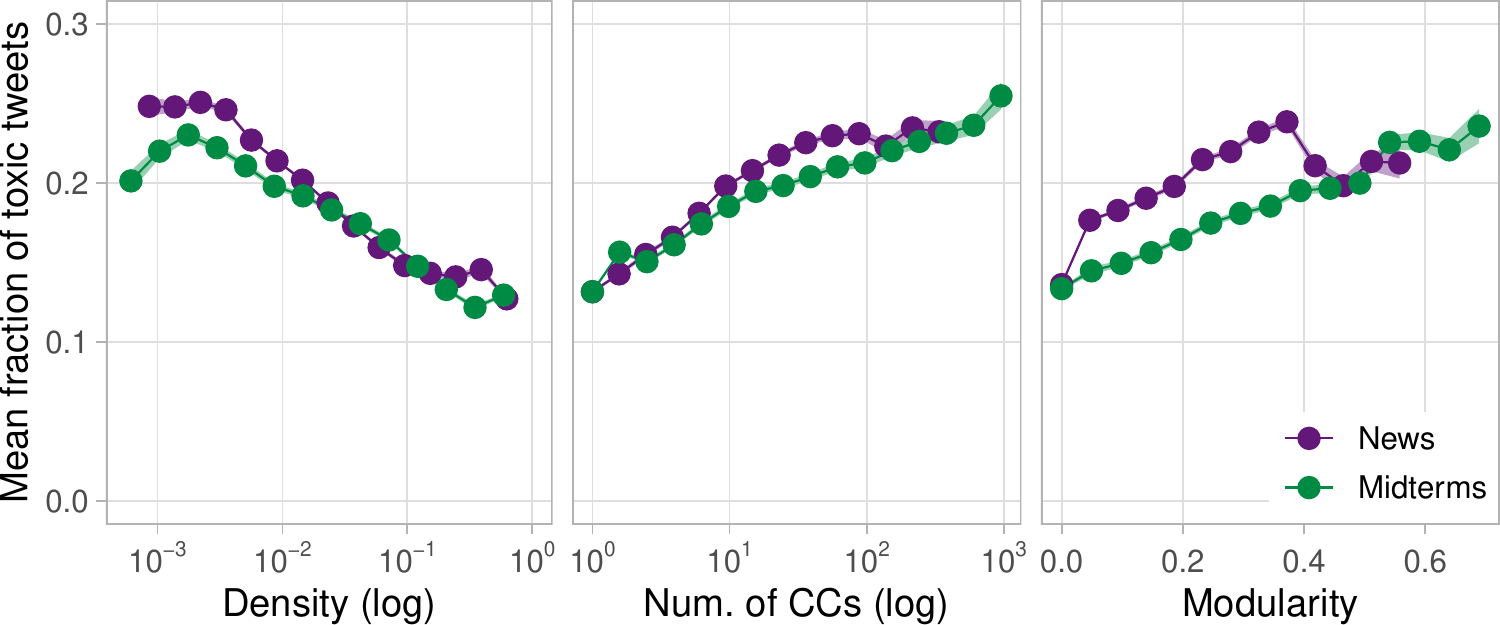}
\caption{Relationship between the density, number of connected components, the modularity of the  conversation follow graph and the mean fraction of toxic tweets.}
\label{fig:analysis-follow-graph}
\end{figure}

\textbf{Graph size and density.}
We start by investigating how the size of the graph is related to the overall toxicity. Unsurprisingly, given our reply tree analysis, larger follow graphs tend to be more toxic. However, we find that the density of the participants' connections also matters (Figure~\ref{fig:analysis-follow-graph}, left). Conversations in which the participants are more densely connected in the follow graph tend to be less toxic in both datasets. Larger follow graph density suggests that the conversation participants are more familiar with each other, which increases the social cost of toxic behavior.

While it is clear that a higher density of connections among the conversation participants correlates negatively with overall toxicity, it is unclear whether the way these connections are distributed in the follow graph impacts toxicity. A follow graph may have high density because there are tightly-knit groups or many uniformly distributed connections among the users. Next, we analyze the number of connected components and modularity of the follow graphs to answer this question.

%
%

\textbf{Number of connected components.}
We start by looking at the relationship between the number of connected components in the follow graph and overall toxicity. A connected component of a graph is a subgraph in which there is a path between any pair of nodes in the subgraph and no path to nodes in the rest of the graph. We compute the weakly connected components of the follow graphs, i.e., we ignore the direction of the edges. The number of connected components has been recently used to quantify the structural diversity of an individual's ego graph and has been shown to explain product adoption decisions made by the ego~\cite{ugander2012structural}. In the context of the conversation follow graph, a larger number of connected components suggests that there are many groups of participants who know each other but do not know any other conversation participants. We find that the number of connected components is positively correlated with the overall toxicity of the conversation, both in the news and the midterms dataset (Figure~\ref{fig:analysis-follow-graph},~middle).

\textbf{Modularity.}
Given a partitioning of a graph, the modularity measures whether there are more or less edges within the partitions than we would expect at random~\cite{newman2004finding}. It takes positive values if there are more edges within the partitions than we would expect by chance, and negative values if there are less. We first partition the conversation follow graphs using the Louvain algorithm~\cite{blondel2008fast}, and then we compute the modularity of the best partitioning. Partitioning the graph with Louvain is a more flexible way of grouping the users than computing the connected components, allowing for some edges between users of different groups.

We find that conversations in which the follow graph has higher modularity among the participants tend to be more toxic (Figure~\ref{fig:analysis-follow-graph}, right). This pattern holds in both datasets, but it is more pronounced in the midterms dataset. We note that due to the sparsity of many follow graphs, a large fraction of conversations have a modularity value of zero, 69.1\% in the news and 76.35\% in the midterms dataset.

%
%
\section{Predicting Toxicity}
So far, we have demonstrated that there is a strong correlation between toxicity and various structural measures of the conversation after the conversation has ended. Next, we consider two prediction tasks that will allow us to measure the utility of these structural properties in forecasting toxicity. In the first task, we focus on predicting whether the conversation as a whole will become more or less toxic. In the second task, we focus on predicting the behavior of individual users and whether their next reply will be toxic.

\subsection{Conversation Toxicity Predictions}\label{sec:prefix-predictions}
We start with the first task. Given the initial stages of the conversation, e.g., the first ten replies, we are interested in predicting whether the rest of the conversation will turn more or less toxic than expected. To make predictions, we will compute various metrics that characterize the relations among the tweets and the users in the conversation prefix.

Beyond allowing us to evaluate which metrics are good indicators of future toxicity, this task also has several practical applications. First, accurate predictions of future toxicity can be used to decide how much visibility a conversation should be given. For instance, if we suspect that a conversation will turn very toxic, we may decide to downrank the root tweet in users’ feeds. These predictions can also be combined with engagement predictions to surface relevant but nontoxic conversations. Second, early warnings of derailment can be used to prompt the initiator of the conversation to moderate the discussion and prevent it from turning toxic. This is particularly useful for accounts that post frequently, such as news outlets, but do not have the capacity to monitor the conversations.
Twitter has recently released new features that allow users to actively moderate the conversations prompted by their tweets by hiding some replies.

\begin{table*}[t]
\centering
\caption{Features used in the future conversation toxicity prediction task.}
\small
\begin{tabular}{ p{24mm} p{146mm} }
\toprule
\textbf{Feature Set} & \textbf{Features} \\
\midrule
Content Toxicity &
Mean/std/min/max/quartiles of the toxicity scores of the prefix tweets. \\
Reply Tree &
Depth $\bigcdot$
width $\bigcdot$
Wiener index $\bigcdot$
number of nodes at depth $i$ $\bigcdot$
mean/var/h-idx/gini/entropy of the number of nodes at every depth $\bigcdot$
depth / size ratio $\bigcdot$
mean/var/h-idx/gini/entropy of the depths of all and leaf nodes $\bigcdot$
mean/var/h-idx of number of children per node $\bigcdot$
fraction of direct replies in total and with a reply $\bigcdot$
gini/entropy of subtrees at depth 1 $\bigcdot$
depth / size ratio of the largest subtree $\bigcdot$
assortativity in political alignments $\bigcdot$
mean/var/h-idx/gini/entropy of number of tweets per user. \\
Follow / Reply Graph &
Number of nodes $\bigcdot$
num. of edges $\bigcdot$
density $\bigcdot$
mean/var/fraction-positive/h-idx/gini of node in/out/total degrees $\bigcdot$
degree and in-out degree assortativity $\bigcdot$
number/fraction of pairs of nodes with no/1-way/2-way edges $\bigcdot$
fraction of connected node pairs $\bigcdot$
betweenness/closeness/eigenvalue/pagerank centralization $\bigcdot$
algebraic connectivity of the largest CC $\bigcdot$
local/global clustering coefficient $\bigcdot$
modularity of the best Louvain partitioning $\bigcdot$
fraction of nodes in the largest CC $\bigcdot$
num. of CC with at least 1/2/3/5/10 nodes $\bigcdot$
num. of nodes/edges/density/CCs in the $k$-core/$k$-truss for $k$=1...5 $\bigcdot$
mean/var/h-idx/gini of number of followers/friends $\bigcdot$
number of friends/followers assortativity $\bigcdot$
political alignment assortativity/modularity. \\
Subgraphs &
Dyadic and triadic census of the follow, reply, and the intersection of the follow and reply graphs. \\
Embeddedness &
Number/mean/var/entropy/gini of the number/fraction of common friends among all pairs of users that have no/1-way/2-way connections in overall, in the follow graph, in the reply graph. \\
Political Alignment &
Mean/std/min/max/quartiles/IQR of users' political alignments $\bigcdot$
num. left/num. right/entropy of the users' political leanings. \\
Arrival Sequence &
Temporal id (assigned sequentially to the every new replier) of the $i$th reply's poster $\bigcdot$
number of unique users up to the $i$th reply. \\
Rate &
Time between the root tweet and the $i$th reply, time between reply $i-1$ and $i$ $\bigcdot$
mean time between all replies, replies in the first and second half of the conversation prefix. \\
\bottomrule
\end{tabular}
\label{tab:prefix-features}
\end{table*}

\textbf{Controlling for prefix toxicity.}
A common way to formulate the task for the prediction problem we are interested in is to predict whether the level of toxicity in the conversation suffix will be above or below the median toxicity of all conversations. For instance, this setup has been used to predict whether a conversation thread will grow~\cite{backstrom2013characterizing} or whether an information cascade will grow~\cite{cheng2014can}. However, our scenario is slightly different as the toxicity in the suffix is confounded by the toxicity in the prefix. Even if we fix the prefix size, different conversations may contain a different number of toxic tweets in the prefix.
Comparing the relationship between the toxicity in the prefix (i.e., in the first $k$ tweets) and in the suffix (i.e., the rest of the conversation), we find that conversations with more toxicity in the prefix tend to have a higher fraction of toxic tweets in the suffix.

To address this issue, for each prefix size, we first bucket the conversations by the number of toxic tweets in the prefix and then assign the labels depending on whether the fraction of toxic tweets in the suffix is above or below the median of all conversations in the bucket. For example, given the first ten replies of the conversation, four of which are toxic, we aim to predict whether the toxicity in the conversation suffix will be higher than the median toxicity of all conversations in the training set with four toxic tweets within the first ten tweets. To ensure that there are enough positive and negative examples in each bucket, we only consider buckets with at least 200 conversations. We also exclude conversations smaller than twice the prefix size to ensure that we have a reasonable estimate of the fraction of toxic tweets in the suffix. This process results in a balanced dataset in which there is no correlation between the labels and the number of toxic tweets in the prefix.

\textbf{Methods used for learning.}
We tested a variety of linear and non-linear machine learning methods, including Logistic Regression, Linear SVM, Random Forests, and Gradient Boosted Regression Trees (GBRTs). We find that non-linear models perform significantly better (with increases in accuracy between 2\% to 5\%) and that among them, GBRTs perform best. To simplify the exposition of the results, we only report the performance of the GBRT models.

To evaluate the performance of the models, we used nested cross-validation: in the inner-loop, we perform 5-fold cross-validation to select the best hyper-parameters and refit the model with the best settings, and in the outer-loop, we perform 10-fold cross-validation to measure the performance of the tuned model on unseen data. This procedure leads to unbiased estimates of the expected accuracy of the models after hyper-parameter tuning~\cite{cawley2010over}. We tuned only one hyper-parameter, the number of GBRT estimators, choosing one of the following values: $n_{estimators} \in \{$10, 25, 50, 100, 500, 1k, 2k, 3k, 5k, 10k$\}$. We report the mean and 95\% confidence intervals of the classification accuracy computed across the 10 outer folds.

\begin{figure*}
\centering
\includegraphics[width=\linewidth]{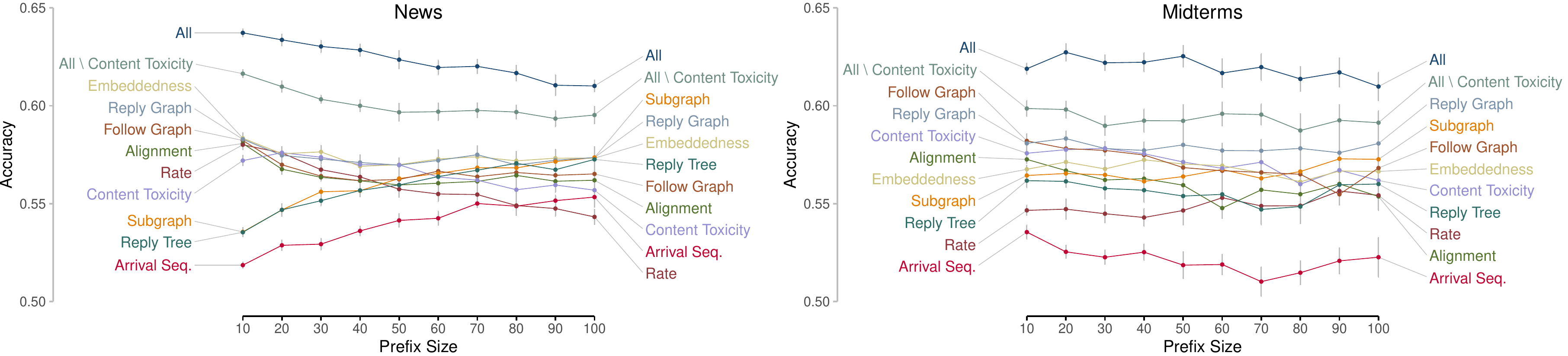}
\caption{Results of the future conversation toxicity prediction task for different prefix sizes in both datasets.}
\label{fig:prefix-results}
\end{figure*}

\textbf{Feature sets.}
Next, we describe the features that we use to predict future toxicity. These features aim to characterize the relationships between the tweets and the users in the initial stages of the conversations. We group the feature into nine feature sets. We briefly describe each feature set below, and we provide a detailed list of all features in Table~\ref{tab:prefix-features}.

\mybullet
\textit{Content Toxicity Features}: to test how predictive the content of the tweets in the prefix is, we compute summary statistics of the Perspective API toxicity scores. Note that we control for the number of toxic tweets in the prefix after binarizing the toxicity scores, which leads to some loss of information from the toxicity scores; e.g., two conversations may have the same number of toxic tweets in the prefix, but the tweets in one of them may have much higher toxicity scores. Prior work has shown that this approach performs significantly better than bag-of-words and sentiment analysis, and on par with hand-crafted linguistic features designed for this task~\cite{zhang2018conversations}.

\mybullet
\textit{Reply Tree Features}: characterize various aspects of the reply tree structure. We significantly expand on the features from \S \ref{sec:analyses-reply-tree}.

\mybullet
\textit{Follow / Reply Graph Features}: capture the overall conversational and social structure of the conversation. We compute various statistics of both the directed and undirected versions of these graphs, including size, density, degree distribution, assortativity, centralization, clustering, modularity, and structural diversity.

\mybullet
\textit{Subgraph Features}: further characterize the structure of the follow and the reply graphs, as well as their intersection. Previous work has shown that they are useful in detecting controversial conversations~\cite{coletto2017automatic}.

\mybullet
\textit{Embeddedness Features}: measure the overlap of the social contexts/interests among the conversation participants. They allow us to go beyond the direct connections and capture broader, more contextual information about the participants' relationships.

\mybullet
\textit{Political Alignment Features}: capture the overall political alignment of the conversation participants. To measure each user's political alignment, we average the alignment scores of the URL domains~\cite{bakshy2015exposure} the user shared. To infer the user's political leaning, we threshold their political alignment at zero.

\mybullet
\textit{Arrival Sequence Features}: summarize the specific order in which users contribute to the conversation. Previous work has demonstrated that these features are predictive of conversation growth~\cite{backstrom2013characterizing}.

\mybullet
\textit{Rate Features}: encode the ``speed'' at which the initial stage of the conversation unfolded. They have been shown to be indicative of future growth of both conversations~\cite{backstrom2013characterizing} and information cascades~\cite{cheng2014can} on Facebook.

\textbf{Results.}
We evaluate the models’ performance in predicting future toxicity, given different conversation prefix sizes. Due to space constraints, we report only the classification accuracy over the 10 (outer) cross-validation folds; the AUC and F1 scores lead to the same substantive conclusions. Since the datasets are balanced, random guessing would result in a performance of 0.5.

We find that combining all features leads to the best prediction performance with accuracy ranging between 0.61 and 0.64 in the news and between 0.61 and 0.63 in the midterms dataset (Figure~\ref{fig:prefix-results}). While each feature set is individually significantly better than predicting at random, it is the reply graph and embeddedness feature sets that perform best across different prefix sizes in the news dataset, and the reply graph feature set in the midterms dataset.

To measure the contribution of the content toxicity features to the performance of the full model, we train a classifier with all but the content toxicity features (All \textbackslash ~ Content Toxicity). This is the only feature set that relies on the content of the conversation. We find that combining all structural features (All \textbackslash ~ Content Toxicity) leads to significantly better accuracy than using the content toxicity features. However, combining the structural and content toxicity features (All) significantly improves the models' overall performance. This indicates that these two types of features capture different and complementary aspects of the conversation prefix.

Both the rate and arrival sequence features perform better than random. In the news dataset, the rate features are more predictive of toxicity in the initial stages, while the arrival sequence features are more predictive in the later stages of the conversations. In the midterms dataset, the rate features are more predictive than the arrival sequence features. However, both feature sets perform worse than most of the other structural features. This suggests that the structural characteristics that predict conversation toxicity are different from those that predict the growth of conversations~\cite{backstrom2013characterizing} and information cascades~\cite{cheng2014can}.

Intuitively, we would expect the performance of the classifiers to increase as we observe more of the conversation. However, we find the opposite to be true, especially in the news dataset. We offer two possible explanations for this phenomenon. First, as we increase the prefix size, the number of conversations that are big enough to be considered decreases significantly. For instance, there are 149k conversations in the news prefix 10 dataset and 36k data points in the news prefix 100 dataset. Less training data makes generalization harder and often results in lower prediction performance on unseen data. Second, since we define the classification labels by bucketing the conversations by the number of toxic tweets in the prefix, the prediction problem itself becomes harder and more nuanced as we increase the prefix. There are significantly more prefix toxicity buckets as we increase the size of the prefix. In fact, due to the differences in the distribution of conversation sizes between the two datasets, for larger prefix sizes there are fewer prefix toxicity buckets that meet the minimum threshold in the midterms than in the news dataset. This may explain the steeper decrease in performance for larger prefix sizes in the news dataset.

Finally, to contextualize the prediction performance and understand how early the models can warn of conversation toxicity, we compute the median time it takes for a conversation to reach a certain size. The conversations in the news dataset grow much faster than those in the midterms dataset. This is not surprising given the much higher follower counts of the news outlets. In the news dataset, half of the conversations have 10 replies within the first 5 minutes and reach a size of 100 within 30 minutes. In the midterms dataset, half of the conversations reach size of 10 within an hour and size of 100 within 130 minutes. This suggests that we can give a reasonably accurate warning that the conversation may become toxic as early as 5 minutes after the root tweet was posted in the news dataset and within one hour in the midterms dataset.

%
%
\subsection{Next Reply Toxicity Predictions} \label{sec:next-reply-predictions}
In the first prediction task, we aimed to predict how the conversation, as a whole, would unfold in the future by characterizing how the participants are connected to and interact with each other during the initial stages of the conversation. In the second prediction task, we focus on forecasting the behavior of individual users.

\begin{table*}[t]
\centering
\caption{Features used in the next reply prediction task.}
\small
\begin{tabular}{ p{28mm} p{143mm} }
\toprule
\textbf{Feature Set} & \textbf{Features} \\
\midrule
Conversation State &
Number of replies, number/fraction of toxic replies in overall, from and to the user in the conversation so far. \\
User-Parent / User-Root Dyads &
Is parent/root tweet toxic? $\bigcdot$
follow edge type (\lf/\rf/\mf/\nf) $\bigcdot$
number/fraction of common friends $\bigcdot$
$\Delta$ in friend/follower counts $\bigcdot$
$\Delta$ in degree/betweenness/closeness/eigenvalue/pagerank centrality scores in the directed/undirected version of \textit{follow} \& \textit{reply} graphs [$\ast$] $\bigcdot$
same CC/Louvain partition in the \textit{follow} \& \textit{reply}? [$\ast$] $\bigcdot$
$\Delta$ in CC/Louvain partition size in the \textit{follow} \& \textit{reply}~[$\ast$] $\bigcdot$
fraction of toxic replies between the user and the parent/root $\bigcdot$
number of replies/number and fraction of toxic replies user$\rightarrow$parent/root and parent/root$\rightarrow$user $\bigcdot$
$\Delta$ in political alignment $\bigcdot$
same political leaning? \\
Follow / Reply Graph &
Degree/betweenness/closeness/eigenvalue/page-rank centrality of the user in directed/undirected version of graph $\bigcdot$
number/fraction of in/out/2-way edges between the user and other toxic/nontoxic users in the directed/undirected graph $\bigcdot$
number/fraction of other nodes in the same CC/Louvain partition as the user $\bigcdot$
n/mean/var/entropy/gini of the number/fraction of common friends with other users who are not connected to the user/the user is connected to/are connected to the user. \\
Reply Tree &
Tweet depth $\bigcdot$
number of siblings $\bigcdot$
size/fraction of tweets of the subtree the reply belongs to $\bigcdot$
subtree size-depth ratio. \\
Overall Embeddedness &
n/mean/var/ent/gini of user's number/fraction of common friends with all other conversation participants. \\
Toxic Embeddedness &
n/mean/var/ent/gini of user's number/fraction of common friends with users with \textit{at least one toxic tweet}/\textit{no toxic tweets}. \\
Political Alignment &
Mean $\Delta$ between the user’s alignment score and all other users $\bigcdot$
fraction of other users with the same political leaning. \\
User Info &
Number of friends/followers in the follow graph $\bigcdot$
friends-followers ratio. \\
\bottomrule
\multicolumn{2}{r}{Note: [$\ast$] = only a user-parent dyad feature}
\end{tabular}
\label{tab:next-reply-features}
\end{table*}

In particular, we aim to predict whether the next reply by a specific user will be toxic, given the conversation so far and the user’s relationship to other conversation participants. This prediction problem is inspired by the practical need to rank the different branches of a conversation to present them to the end-user in a linear order (Figure~\ref{fig:dtox-hero}A). While Twitter conversations have a tree structure (Figure~\ref{fig:dtox-hero}B), Twitter’s user interface displays the replies in a linear order, which requires one to decide how to order the different branches of the conversation tree. If we can estimate how likely the user is to post a toxic reply to each of the conversation branches, then we can display the branches for which the user is least likely to post a toxic reply first. This will make it less likely for the user to reach parts of the conversation that may prompt them to post a toxic reply.

Unlike the previous prediction problem where we did not know who will contribute to the conversation next, here we assume that we know the identity of the user who will reply next, but we do not know whether their reply will be toxic or not. This setup matches exactly the scenario that we would face in a production system: when a specific user opens a tweet, we need to decide how to rank the reply branches of the conversation such that, if they post a reply, they are more likely to post a nontoxic one. Moreover, this setup creates an opportunity for building personalized models that rank the branches of the conversation based on the identity of the user viewing the conversation.

\textbf{Controlling for content.}
The content of the root tweet may, to a large extent, drive the structure and the toxicity of the conversation. For instance, tweets by news outlets that cover divisive topics or tweets by midterm candidates sharing their policies on contested issues may be more likely to spur toxic conversations. Moreover, the content discussed across different communities (e.g., audiences of different news outlets) may vary significantly. These considerations motivate the need for an experimental setup that allows us to evaluate the predictive power of the metrics that we propose, but factors out the influence of the content.

We control for the content by using a paired prediction scheme: for each conversation, we sample a toxic and nontoxic tweet pair and aim to predict which one of the two tweets is more likely to be toxic. Each pair of tweets is one instance of the prediction task. For each tweet, we reconstruct the conversation up to the point before it was posted. To represent a pair, we take the difference of the feature vectors of the individual tweets and define the label as positive if the first tweet was toxic and negative otherwise. To ensure a balance between the positive and negative class, we construct the pairs such that in exactly half of them, the first tweet is toxic. To avoid overrepresenting any one conversation, we sample at most one pair per conversation. While sampling tweets, we exclude self-replies and direct replies to the root as we are interested in indicators of toxicity among the conversation participants. We also exclude tweets whose Perspective API toxicity scores were close to the threshold, considering only tweets with scores below~0.25~or~above~0.75.

This paired prediction scheme has been used to control for content in several previous studies~\cite{zhang2018characterizing, chang2019trouble}. While controlling for the content makes the prediction task harder, it allows us to measure the predictive power of the conversations' structural representation.

\textbf{Methods used for learning.} As in the previous task, we use GBRTs and a nested 10-fold cross-validation setup.

\textbf{Feature sets.}
We consider features that capture the structural relationship between the next reply and tweets in the conversation so far, and between the user and the current conversation participants. We group the features into ten feature sets (full list in Table~\ref{tab:next-reply-features}):

\mybullet
\textit{Conversation State Features:} capture the size and toxicity of the conversation before the next reply was posted, including the number of tweets, number of toxic tweets overall, and posted by the focal user or posted in reply to previous tweets by the focal user. Note that these features primarily rely on the toxicity of the tweets'~content.

\mybullet
\textit{User-Parent / User-Root Dyad Features}: encode the various aspects of the relationship and previous interactions between the user and the parent (i.e., the participant they are replying to) / the root (i.e., the user that started the conversation).

\mybullet
\textit{Reply Tree Features}: capture the position of the eventual reply in the reply tree.

\mybullet
\textit{Follow / Reply Graph Features}: characterize the user's position in the two graphs, the number of edges to toxic (i.e., users with at least one toxic tweet) and nontoxic conversation participants, and the number of common friends with other conversation participants who are connected vs. not connected to the user.

\mybullet
\textit{Overall / Toxic Embeddedness Features}: measure the overlap between the social context within the Twitter follow graph of the user and all other conversation participants / participants that posted at least one vs. no toxic tweets.

\mybullet
\textit{Political Alignment Features}: encode how similar is the political alignment/leaning of the user to the alignment/leaning of the other conversation participants. We measure political alignment/leaning as described in the previous task.

\mybullet
\textit{User Information Features}: capture general user information.

\textbf{Results.}
To evaluate the features' predictive power, we sample 96,520 pairs of tweets from the news and 50,143 pairs of tweets from the midterms dataset, sampling each pair from a different conversation. We report the classification accuracy, AUC, and F1 scores over 10 (outer) cross-validation folds. Since the datasets are balanced, random guessing would result in a performance of 0.5.

\begin{figure}
\centering
\includegraphics[width=\linewidth]{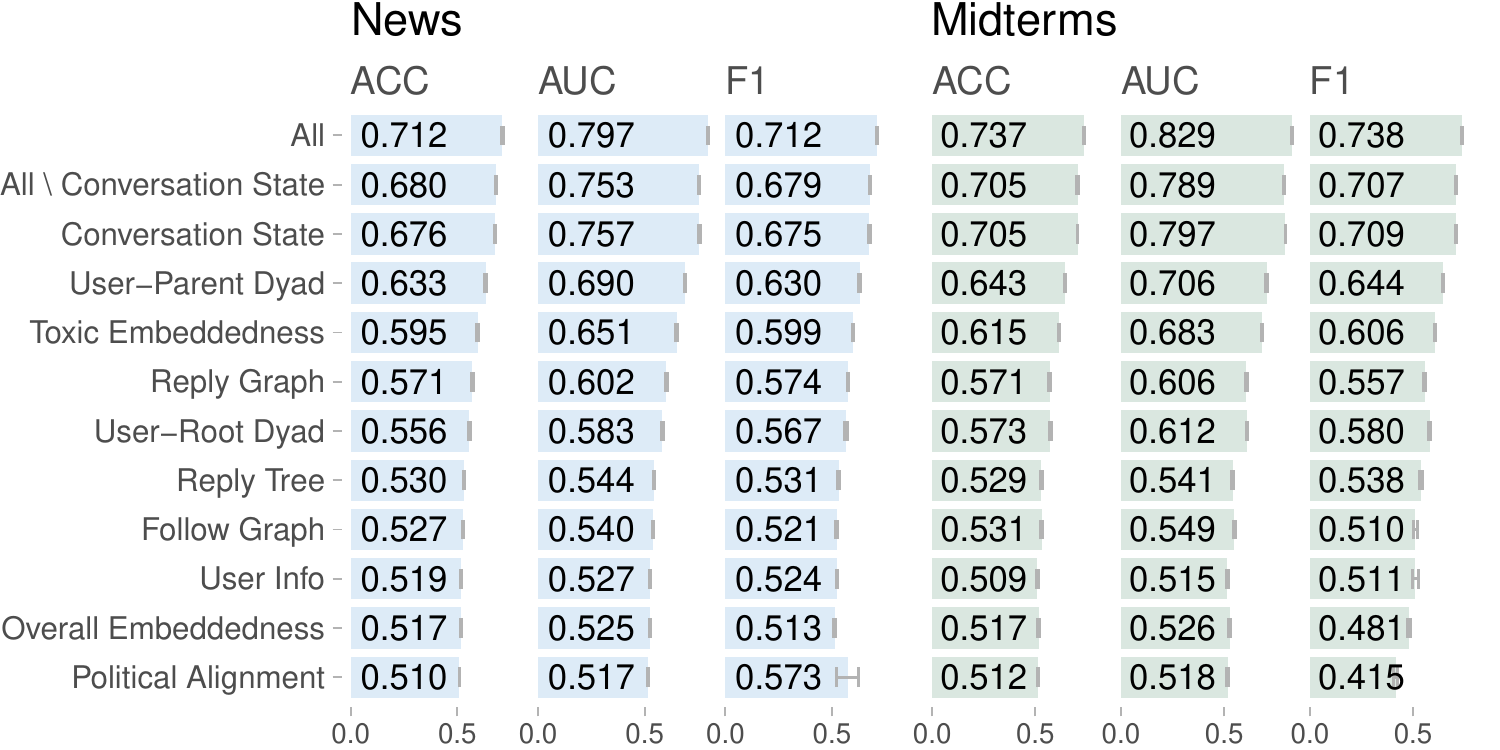}
\caption{Results for the next reply prediction task.}
\label{fig:next-reply-results}
\end{figure}

We find that when we use all features, the models achieve surprisingly strong performance: accuracy of 0.712 and AUC of 0.797 in the news dataset, and accuracy of 0.737 and AUC of 0.829 in the midterms dataset (Figure~\ref{fig:next-reply-results}). When we consider the performance of the individual feature sets, we observe that features capturing the conversation state perform best. To investigate whether the models would perform well without the conversation state features, which primarily rely on the content, we trained a classifier which excluded them. We find that while the classification accuracy drops by 0.032 in both datasets, we still obtain good performance even without these features. We achieve similar performance when we use just the conversation state features and just the structural features (Figure~\ref{fig:next-reply-results}, All \textbackslash ~Conversation State). Moreover, we find that combining the two (i.e., using all features) significantly improves the classification performance in both datasets. This indicates that the conversation state features and the structural features capture different and complementary aspects of the conversation that are predictive of whether the next reply will be toxic.

We find that both the absolute and relative performance of the individual features sets is similar in the two datasets. This is perhaps because our experimental setup was designed to control for the content of the root tweets and suggests that the proposed features could generalize beyond these two datasets. We also observe that most feature sets perform significantly better than random, which indicates that the predictions of the full model do not rely on any individual feature set and demonstrates the models' robustness.

\subsection{Domain Transfer}
To test how the models generalize across domains, we split each dataset into 80\% train and 20\% test set, fit a model on the training set of one dataset (tuning the hyper-parameters with cross-validation), then test on its own test set (e.g., news on news) and the test set of the other dataset (e.g., news on midterms). On the future conversation toxicity prediction task (\S \ref{sec:prefix-predictions}), we find that the accuracy on the other test set drops by 1.7\%--6.2\% in the news and 1.7\%--4.7\% in the midterms dataset---in both cases, the models still perform much better than random. On the next reply prediction task (\S \ref{sec:next-reply-predictions}), the accuracy drops by $<$1\% in both datasets. These results suggest that the models in both tasks make sensible out-of-domain predictions.

\section{Further Related work} \label{sec:related-work}
In addition to the literature on conversation analysis previously mentioned, our work builds on and contributes to a rich body of prior work on antisocial behavior online.

Recent studies have sought to analyze various aspects of antisocial users, such as their activity patterns~\cite{chatzakou2017mean, ribeiro2018characterizing}, their evolution from the time they join a community until they become toxic~\cite{cheng2015antisocial}, and the primary triggers of their behavior~\cite{cheng2017anyone}. Another line of work has focused on detecting antisocial behavior by analyzing the relationship between the instigators and their targets, including their linguistic similarities~\cite{liu2018forecasting, ziems2020aggressive}, their shared social context~\cite{radfar2020characterizing}, their personalities~\cite{elsherief2018peer}, and the misalignment between their intentions and perceptions~\cite{chang2020don}. Other studies have explored antisocial behavior at the community level, including inter-community conflict~\cite{kumar2018community}, the maintenance of toxic community norms~\cite{rajadesingan2020quick}, and the effects of major negative events~\cite{lamba2015tempest}. In contrast to these studies, in this paper, we focus on the conversational and social context in which toxic behavior occurs.

Most related to the current study is the work by Zhang et al.~\cite{zhang2018characterizing}, which aims to predict whether the conversation participants will block each other using a hypergraph representation of their prior interactions in the conversation. In this work, we focus on predicting toxic behavior, which is substantially more prevalent than blocking, and consider both the conversational (through the reply tree and the user reply graph) and the social (through the follow graph) structure of the conversations.

Finally, several recent studies have examined Twitter interactions in the context of the 2018 US midterm elections but have focused on the users who engage in adversarial interactions against the candidates~\cite{hua2020towards, hua2020characterizing}. In this work, we focus on the conversations that are started by or mention the candidates and study how the structural characteristics of the conversations are related to toxicity.

\section{Discussion and Conclusion}
In this paper, we focused on the structural view of conversations on Twitter and the relationship to toxicity. We examined 1.18M conversations rooted in tweets that are posted by or mention the accounts of major news outlets and 2018 midterm election candidates.

We analyzed the conversations at the individual, dyad, and group levels. At the individual level, we found that toxicity is spread among many low to moderately toxic users (\S \ref{subsec:individual-level}). At the dyad level, we found that toxic posts are more likely to receive toxic replies and that toxic replies are more likely to come from users who do not have a social connection with the poster (\S \ref{sec:dyads}). At the group level, we found that toxic conversations tend to be larger, have wider and deeper reply trees, but less dense follow graphs. (\S \ref{sec:analyses-reply-tree} \& \ref{sec:analyses-follow-graph}).

To test the utility of the structural features of the conversations in forecasting toxicity, we considered two prediction tasks. In the first task, we showed that we can predict with an accuracy of up to 61.6\% whether the conversation will become more toxic than expected, given the first ten replies and using only the structural features (\S \ref{sec:prefix-predictions}). In the second task, we demonstrated that the structural features can also predict whether the next reply posted by a specific user will be toxic, with an accuracy of up to 70.5\% (\S \ref{sec:next-reply-predictions}).

These findings advance our understanding of the social conditions that lead to toxic behavior online and inform the design of healthier social media platforms. In particular, we suspect that the behavior of many low and moderately toxic users is modulated by their awareness of the social context of the conversation, their position within it, and who will observe their behavior. This suggests follow-up experiments in how social connections could be exposed to potentially drive toward more civil behavior.
Our work also has important practical implications. The predictions of future toxicity based on the initial replies can be used as a signal to rank conversations at their early stages or to alert the user who initiated the conversation that it may derail and prompt them to intervene. The predictions of whether the next reply posted by a specific user will be toxic can be used to rank the different threads of the conversation such that the user is less likely to post a toxic~reply.

Nonetheless, our work has limitations that open new avenues for future work.
First, since detecting toxic content was not the primary goal of this study, we relied on the Perspective API and adopted a broad definition of toxicity. Several recent studies have identified more granular types of antisocial behavior~\cite{founta2018large}. One direction for future work is to examine how the conversational structure is related to different types of antisocial behaviors. Second, while we analyzed conversations prompted by a wide range of accounts, we mostly focused on political conversations. Future work may examine how our findings generalize to other domains.
Third, we focused on finding associations and testing the predictive power of the conversation structure. The natural next step is to run A/B tests that examine whether our findings can be used to reduce toxicity.
Last, we based our analysis on Twitter, but our methodology applies to other social platforms that allow users to engage in conversations and friend/subscribe/follow each other, such as Facebook, Youtube, and Reddit. While we suspect that many of our findings will hold, investigating the idiosyncrasies of different platforms is an important area for future research.

\textbf{Acknowledgements.} We are grateful for comments from Lada Adamic, Dean Eckles, Johan Ugander, Kiran Garimella, Peter Beshai, members of the Twitter HUB team, Facebook CDS team, Social Analytics Lab at MIT, Ugander Lab at Stanford, and the anonymous reviewers. We thank Twitter for the financial and data support.

\bibliographystyle{ACM-Reference-Format}
\bibliography{refs}

\end{document}